\documentstyle[prl,aps,epsf,epsfig,multicol]{revtex}

\begin{document}

\title{A New Monte Carlo Algorithm for Protein Folding} 

\author{Helge Frauenkron$^{1)}$, Ugo Bastolla$^{1)}$, Erwin Gerstner$^{1), 2)}$, 
      Peter Grassberger$^{1), 2)}$, and Walter Nadler$^{1)}$}

\address{$^{1)}$ HLRZ c/o Forschungszentrum J\"ulich, D-52425 J\"ulich, 
Germany, and\\
$^{2)}$ Physics Department, University of Wuppertal, D-42097 Wuppertal, Germany}

\date{\today}

\maketitle

\begin{abstract}
  We demonstrate that the recently introduced pruned-enriched Rosenbluth 
  method (PERM) leads to extremely efficient algorithms for 
  the folding of simple model proteins. We test them on several 
  models for lattice heteropolymers, and compare to published Monte 
  Carlo studies. In all cases our algorithms are faster than 
  previous ones, and in several cases we find new minimal energy 
  states. In addition, our algorithms give estimates 
  for the partition sum at finite temperatures.

  \vspace{4pt}
\end{abstract}

\begin{multicols}{2}

Protein folding \cite{creighton} is one of the outstanding problems in 
mathematical biology. It is concerned with the problem of 
how a given sequence 
of amino acids assumes precisely that geometrical shape which is 
biologically useful. Currently it is much easier to find coding DNA (and, thus,
amino acid) sequences than to find the corresponding structures. 
Thus, solving 
the folding problem would be a major break-through in understanding 
the biochemistry of the cell, and in designing artificial proteins. 

In this Letter we are concerned only with the most straightforward direct 
approach: given a sequence, a molecular potential and no other information, 
find the ground state and the equilibrium state at physiological 
temperatures. Note that we are not concerned with the kinetics of 
folding, but only in the final outcome. Also, we will not address 
the problems of how to find good molecular potentials, and what is 
the proper level of detail in describing proteins. Instead, we will 
use simple coarse-grained models which have been proposed in the 
literature and have become standards in testing the efficiency of 
folding algorithms. 

The models we study are heteropolymers which live on 3-$d$ 
or 2-$d$ regular lattices. They are self-avoiding chains with attractive 
or repulsive interactions between neighboring non-bonded monomers. These 
interactions can have continuous distributions \cite{klimov}, but the 
majority of authors considered only two kinds of monomers. 
In the HP model \cite{dill85,dill89-92} 
they are hydrophobic (H) and polar (P), with 
$(\epsilon_{HH},\epsilon_{HP},\epsilon_{PP})=-(1,0,0)$. 
Since this leads to highly degenerate 
ground states, alternative models were proposed, e.g. 
$\vec\epsilon=-(3,1,3)$ \cite{socci1} and  
$\vec\epsilon=-(1,0,1)$ \cite{otoole}. 

The algorithms we apply here are variants of the pruned-enriched Rosenbluth 
method (PERM) \cite{alg}, a chain growth algorithm based on 
the Rosenbluth-Rosenbluth (RR) \cite{rr} method. Monomers 
are added sequentially, the $n$-th monomer being placed at site $i$
with probability $p_n(i)$. In {\it simple sampling}, $p_n(i)$ is
uniform on all neighbors of the last monomer, leading to exponential 
attrition. The original RR method avoids this by using a uniform 
$p_n(i)$ on all {\it vacant} neighbors of $i_{n-1}$. More generally, we 
call any non-uniform choice of $p_n(i)$ a generalized RR method.
The relative thermal weight of a particular 
chain conformation of length $n$ is then determined by
$ W_n =  m_n \exp(-\beta\Delta E_n) W_{n-1}$, with $W_1=1$; 
$\Delta E_n$ is the energy gain from adding monomer $n$; and 
$m_n$ is the Rosenbluth factor, 
$m_n=\sum_{j \in \left\{\rm nn\right\} } p_n(j)/p_n(i)$.
We note that $W_n$ is also an estimate for the partition function $Z_n$
of the $n$-monomer chain \cite{alg}. 
Chain growth is stopped when the final size is reached and started new
from $n=1$.

In easy cases, $p_n(i)$ can be chosen so that Boltzmann and Rosenbluth 
factors -- or Rosenbluth factors for different $n$ -- cancel, leading to 
narrow weight distributions.
But in general, this algorithm produces a wide spread in 
weights that can lead to serious problems \cite{kremer}. On the other 
hand, since the weights accumulate as the chain grows, one can interfere 
during the growth process by `pruning' conformations with low weights 
and enriching high-weight conformations. This is, in principle, 
similar to population based methods in polymer simulations 
\cite{garel,velicson} and in quantum Monte Carlo (MC) \cite{umrigar}. 
However, our implementation is different. Pruning is done stochastically: 
if the weight of a conformation has decreased below a threshold $W_n^<$, 
it is eliminated with probability 1/2, while it is kept and its weight is 
doubled in the other half of cases. Enrichment\cite{enrich} is 
done independently of this: 
if $W_n$ increases above another threshold $W_n^>$, 
the conformation is replaced by $n_c$ copies, each with weight $W_n/n_c$. 
Technically, this is done by putting onto a stack all information 
about conformations which still have to be copied. This is most easily 
implemented by recursive function calls \cite{alg}. Thereby the need for 
keeping large populations of conformations \cite{garel,velicson,umrigar}
is avoided. PERM has proven
extremely efficient for studies of lattice homopolymers near the $\theta$ 
point \cite{alg}, their phase equilibria \cite{multic}, 
and of the ordering transition in semi-stiff polymers~\cite{stiff}. 

The main freedom when applying PERM consists in the a priori choice of 
the sites where to place the next monomer, i.e. the probabilities $p_n(j)$,
in the thresholds $W_n^<$ and $W_n^>$ for pruning and enrichment, 
and in the number of copies, $n_c$,  made upon enrichment. 
All these features do not effect the
correctness of the algorithm, but they can greatly influence its 
efficiency. They may depend arbitrarily on chain lengths and on 
local conformations, and they can be changed freely at any time during 
the simulation. Thus, the algorithm can `learn' during the simulation \cite{alg}.

In order to apply PERM to heteropolymers at very low temperatures, 
the strategies proposed in \cite{alg} are modified as follows.

(1) For homopolymers near the theta-point 
it had been found that the best choice for the placement of 
monomers was not according to their Boltzmann weights, but uniformly on 
all allowed sites \cite{alg,multic} like in the original RR. 
This is due to cancellation between Boltzmann and RR factors: larger 
Boltzmann factors correspond 
to higher densities and, thus, to smaller RR factors \cite{kremer}. 

For a heteropolymer this has to be modified, as there is no longer a 
unique relationship between density and Boltzmann factor. In a strategy 
of `anticipated importance sampling' we should preferentially 
place monomers in sites with mostly attractive neighbors. Assume that we 
have two kinds of monomers and we want to place a type-$A$ monomer. If 
an allowed site has $m_B$ neighbors of type $B$ ($B=H,P$), we select 
this site with probability $\propto 1+a_{AH}m_H+a_{AP}m_P$. Here, $a_{AB}$ 
are constants with $a_{AB}>0$ for $\epsilon_{AB}<0$ and vice versa. 

(2) Most naturally, the $W_n^>$ and $W_n^<$ are 
chosen proportional to the estimated partition sum $Z_n$
(i.e. the average of the $W_n$ already generated) : e.g. 
$W_n^< = cZ_n$, $c\approx0.5$, and $W_n^> = rW_n^<$, $r\approx10$ \cite{alg}.  
But this becomes inefficient at very low $T$ since $Z_n$ will 
be underestimated as long as no low-energy state is found. But when this 
finally happens, $W_n^>$ is too small and, thus, too many (correlated) copies are 
produced. This costs CPU time but does not increase the quality of sampling.

This problem could be avoided by increasing $W_n^>$ and $W_n^<$ during particularly 
successful `tours' (a tour is the set of conformations derived  
from a single start \cite{alg}). But then also the average number of long 
chains is decreased in comparison with short ones. To reduce this effect 
and to create a bias towards a sample which is flat in chain length, 
we multiply by some power of $M_n/M_1$, where $M_n$ is the number of 
generated chains of length $n$. With ${\cal N}(n)$ denoting the number 
of chains generated during the current tour we used 
$
   W_n^< = C\;Z_n\; [(1+{\cal N}(n)/M)(M_n+M)/(M_1+M)]^2 
$,
with $C$ a constant of order unity
and $M$ a constant of order $10^4 - 10^5$.

(3) For the number of copies $n_c$ created when $W_n$
surpasses $W_n^>$, a good choice is int$[1+\sqrt{W_n/W_n^>}]$.

(4) In some cases we did not start to grow the chain from one end but 
from a point in the middle. We grew first into one direction, and then 
into the other. 
Results were averaged over all possible starting points.
The idea behind this is that real proteins have 
folding nuclei, and it should be most efficient to start from such a 
nucleus. In some cases this approach was very successful and speeded up 
the ground state search substantially, in others not. 

(5) Special tricks were employed for `compact' configurations filling a
square or a cube \cite{details}.

Let us now discuss our results. Items (a) to (c) concern the original HP 
model \cite{dill85} with 
$\vec\epsilon=-(1,0,0)$.

(a) Ten sequences of length $N=48$ were given in 
\cite{yue-shak}. Each was designed by minimizing the energy of a particular 
target conformation under the constraint of constant composition.
The authors tried to find lowest energy states with a 
heuristic MC method \cite{Fiebig},
and an exact enumeration of low energy states \cite{yue1}
(which cannot be generalized to other models).
The MC method failed in all but one case. Precise CPU times were not quoted. With 
PERM we succeeded to reach ground states in {\it all} cases, average CPU time
per sequence ranging from few seconds to several hours (all times refer to a
SUN ULTRA SPARC, 167 MHz). We verified also that these ground states are 
highly degenerate, and that there are no gaps between ground and first 
excited states. Thus, none of these sequences are good folders, though they 
were designed specifically for this purpose.

(b) In \cite{unger} two versions of a genetic algorithm were used to 
simulate 2-$d$ HP chains of lengths 20 to 64, and compared to other MC 
algorithms. Ground state energies 
were supposed to be known since the chains had been 
specially designed. \hfil In all cases we reached the ground 

\end{multicols}

\vglue-.2cm
\begin{table}
\caption{ Newly found lowest energy states for binary sequences with 
interactions $\vec\epsilon = (\epsilon_{HH},\epsilon_{HP},\epsilon_{PP})$. 
Configurations are encoded as sequences of {\it r}(ight),{\it l}(eft),
{\it u}(p), {\it d}(own), {\it f}(orward), and {\it b}(ackward).}
\begin{tabular}{cccccc} 
   $N$ & $d$ & $\vec\epsilon$ &    sequence   & old $E_{\rm min}$ & Ref. \\ 
       &     &            & example conformation & our $E_{\rm min}$      &       \\ 
       \hline

  60  & 2 & $-(1,0,0)$ & 
  $ P_2H_3PH_8P_3H_{10}PHP_3H_{12}P_4H_6PH_2PHP $ & $-34$ & \cite{unger} \\
 & & & $r_5d_2lul_3dld_2(ru)_2rd_2ldldrdr_2uluru_2rd_2rdldr_2u_3lu_3rd_2rur$ &$-36$ &  \\ \hline

 100  & 2 & $-(1,0,0)$ &
$P_6HPH_2P_5H_3PH_5PH_2P_2(P_2H_2)_2PH_5PH_{10}PH_2PH_7P_{11}H_7P_2HPH_3P_6HPH_2$ & $-44$ & \cite{pekney} \\
 & & & $r_6ur_2u_3rd_5luldl_2drd_2ru_2r_3(rulu)_2urdrd_2ru_3lur_3dld_2rur_5d_3l_5uldl_2d_3ru_2r_3d_3l_2urul$ & $-47$ & \\ \hline

 100  & 2 & $-(1,0,0)$ &
$P_3H_2P_2H_4P_2H_3(PH_2)_3H_2P_8H_6P_2H_6P_9HPH_2PH_{11}P_2H_3PH_2PHP_2HPH_3P_6H_3$ & $-46$ & \cite{pekney} \\
 & & & $ul_2drdl_2u_3ld_4ldrdl_2u_2l_2d_3l_2uru_3r_2u_3rd_3ru_4rul_5dldr_2d_2luldldrdldlu_3lul_2ulur_2dr_2u_3rd_4l$& $-49$ &  \\ \hline

  80  & 3 & $-(1,0,1)$ & 
 ${PH_2P_3(H_3P_2H_3P_3H_2P_3)_3H_4P_4(H_3P_2H_3P_3H_2P_3)H_2}$ & $-94$ & 
 \cite{otoole} \\
 & & & $lbruflbl_2br_2drur_2dldl_3ulfrdr_3urfldl_3ulurur_3drblul_3br_3bl_3dldrdr_3urul_2dlu 
$ & $-98$ &  \\ 
\end{tabular}

\end{table}

\begin{multicols}{2}

\noindent
state energies proposed by the authors in less than 1h 
CPU time, except for the sequence of length 64. For 
that sequence we obtained $E=-39$, while none of the 
algorithms used in \cite{unger} reached energies below $-37$. For
the chain with length 60, we found several states with $E=-36$ although 
the authors had claimed $E\geq-34$ by construction (see Table~I). 

(c) Two 2-$d$ HP chains with $N=100$ were studied in 
\cite{pekney}. The authors claimed that the native conformations 
are compact, fit exactly into a $10\times 10$ square, and have
energies $-44$ and $-46$. These energies were found by a specially 
designed MC algorithm which should be particularly efficient for compact 
conformations. We found non-compact (degenerate) conformations with energies 
$-47$ and $-49$ (see Table~I), while our lowest energy compact states 
(also degenerate) have $E=-46$ and $-47$ \cite{details}.

(d) Sequences with $N=27$ and with continuous interactions were studied 
in \cite{klimov}. Interaction strengths were sampled from Gaussians
and were permuted to obtain good folders. In all cases we could reach
the supposed ground state energies, within less than 1h in the worst 
case. This time the design had been successful, and all sequences showed 
gaps between the ground state and the bulk of low-lying states. These 
gaps were in some cases filled by conformations which were very 
similar to the ground state, so that they could not prevent these 
sequences to be counted as good folders. In no case we found energies 
lower than those quoted in \cite{klimov}. 

(e) Short sequences with which we had neither problems nor 
surprises were given in several papers: 48-mers in \cite{otoole}, 
27-mers in \cite{socci2}, and sequences with $N\leq 36$ in \cite{yue1}.

(f) The most interesting case is a 2-species 80-mer with interactions 
$-(1,0,1)$ studied first in \cite{otoole}. These particular 
interactions were chosen because it 
was hoped that they would lead to compact conformations. Indeed, the 
sequence was specially designed to form a ``four helix bundle" 
which fits perfectly into a $4\times4\times5$ box (see Fig.~1). Its 
energy in this putative native state is $-94$.
Although the authors of \cite{otoole} used highly optimized codes, they 
were not able to recover this state by MC. Instead, they reached only 
$E=-91$. Supposedly, a different state with $E=-94$ was found in 
\cite{pekney}, but Fig.~10 of this paper, which is claimed to 
show this conformation, has a much higher value of $E$. 

Even without much tuning our algorithm gave $E=-94$ after a few 
hours, but it did not stop there. After a number of  
conformations with successively lower energies, the final candidate 
for the native state has $E=-98$. It again has a highly symmetric 
shape, although it does not fit into a $4\times4\times5$ box 
(see Fig.~2). It has twofold degeneracy (the central $2\times2\times2$ 
box in the front of Fig.~2 can be flipped), and both conformations 
were actually found in the simulations. Optimal parameters for the 
ground state search in this model are $\beta=1/kT\approx 2.0$, 
$a_{PP} = a_{HH} \approx 2$, and $a_{HP}\approx -0.13$. With these, 
average CPU times for finding $E=-94$ and $E=-98$ are 
ca. 20 min and 80 hours, respectively \cite{deutsch}.

A surprising result is that the monomers are arranged in four homogeneous
layers in Fig.~2, while they had formed only three layers in the putative 
ground state of Fig.~1. Since the interaction should favor the segregation 
of different type monomers, one might have guessed that a conformation 
with a smaller number of layers should 
\hfil

\begin{figure}
\begin{center}
\epsfbox{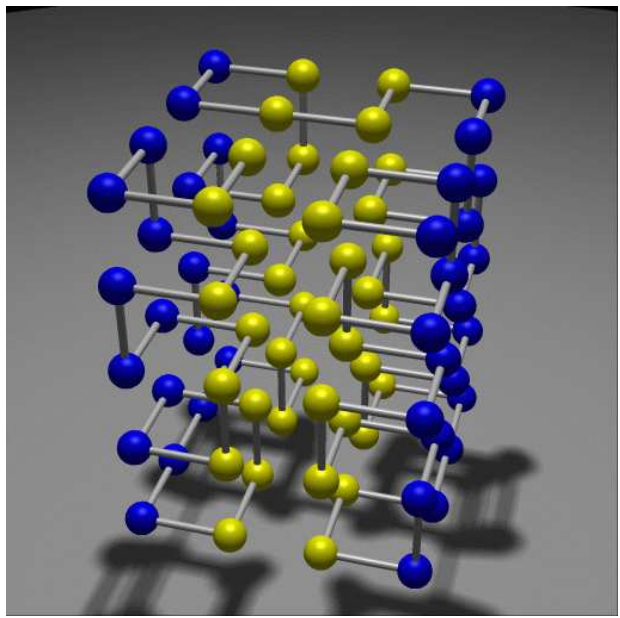}
\vglue0.1cm
\begin{minipage}{8.5cm}
\caption{Putative native state of the ``four helix bundle" 
sequence, as proposed in \protect\cite{otoole}. It has $E=-94$, fits into 
a rectangular box, and consists of three homogeneous layers. Structurally, 
it can be interpreted as four helix bundles.} 
\end{minipage}
\end{center}
\label{ns94}
\end{figure}
\vglue-.5cm

\begin{figure}
\begin{center}
\epsfbox{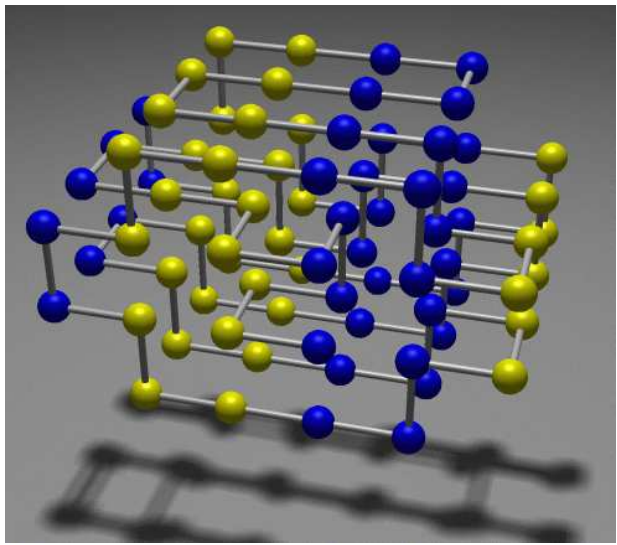}
\vglue0.1cm
\begin{minipage}{8.5cm}
\caption{ Conformation of the ``four helix bundle"
sequence with $E=-98$. We propose that this is the actual ground 
state. Its shape is highly symmetric although it does not fit into a 
rectangular box. It is not degenerate except for a flipping of the central 
front $2\times2\times2$ box.}
\end{minipage}
\end{center}
\label{ns98}
\end{figure}
\vglue-.2cm

\noindent
be favored. We see that this 
is outweighed by the fact that both monomer types can form large double 
layers in the new conformation. Again, our new ground state is not 
`compact' in the sense of minimizing the surface, and hence it disagrees 
with the wide spread prejudice that native states are compact.

\begin{figure}[ht]
\begin{center}\psfig{file=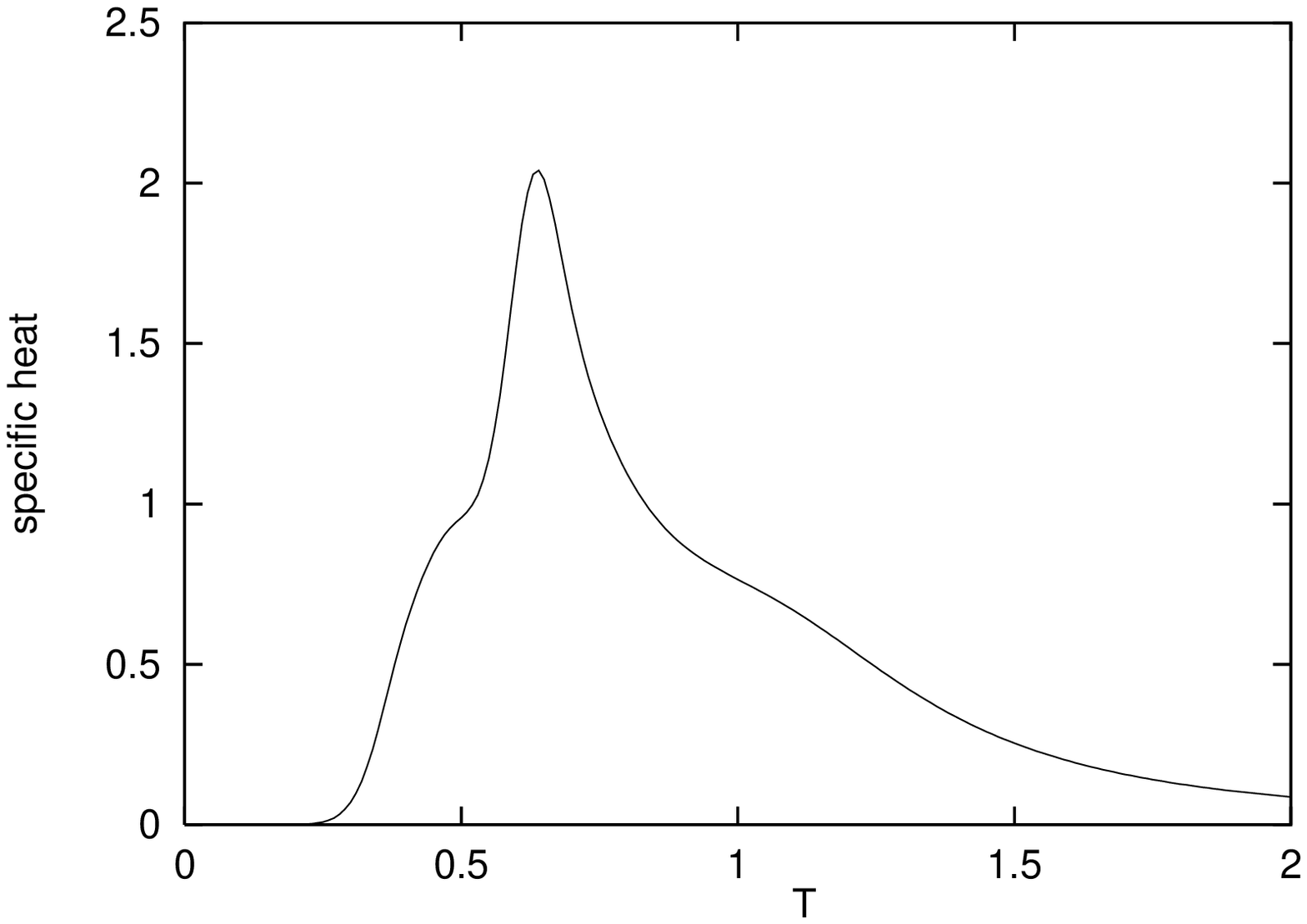,
   width=6.cm,angle=270}\end{center}
{\small FIG. 3: Specific heat (heat capacity per monomer) of the 80-mer 
``four helix bundle" $vs$ $T$.}
\label{cv80}
\end{figure}
\vglue-.3cm

We also constructed histograms of the energy distribution.
Combining them with similar histograms obtained at higher temperatures 
\cite{mchisto} we obtained average energies and heat capacities.
The specific heat (Fig.~3) shows a large peak at $T=0.62$ 
and two shoulders (at 
$T\approx 0.45$ and $1.0$), all of which are statistically significant. 
As shown by a more detailed analysis \cite{details}, the shoulder at 
$T=1$ is due to the collapse from open coil to molten globule, while 
that at $T=0.45$ is due to the folding into the native state. There 
seems to be no state with $E=-97$, and very few states with $E=-96$ 
and $-95$, leading to an effective gap between $E=-94$ and $E=-98$.
The main peak seems related to the formation of -- mostly 
misfolded (helix-dominated) -- secondary and tertiary structure. The 
low-temperature phase, however, contains mostly $\beta$-sheets (see Fig.~2).

In summary, we showed that the pruned-enriched Rosenbluth method can be
very effectively applied to protein structure prediction in simple lattice
models. It is suited for calculating statistical properties and is very
successful in finding native states. In all cases it did better than 
any previous MC method, and in many cases it found lower states than 
those which had previously been conjectured to be native. Especially, we 
have presented a new candidate for the native 
conformation of a ``four helix bundle" sequence which had been studied 
before by several authors. We verified that ground states of the HP 
model are highly degenerate and have no gap, leading to bad folders. 
In contrast, the ground state of the ``four helix bundle" sequence has 
a small gap and has low degeneracy because of the modified interaction 
strengths. But it folds only at very low $T$, and should not 
be a good folder either. 

The authors thank G. Barkema, E. Domany, and M. Vendruscolo
for discussions, and D.K. Klimov and R. Ramakrishnan for correspondence.

\end{multicols}
\end{document}